\RequirePackage{lineno}
\documentclass[prd,twocolumn,showpacs,amsmath,amssymb]{revtex4}
\usepackage{graphicx}
\usepackage{dcolumn}
\usepackage{bm}
\usepackage{psfig}
\usepackage{epsfig}
\usepackage{overpic}
\usepackage{verbatim}
\usepackage[colorlinks,linkcolor=blue,anchorcolor=red,citecolor=blue]{hyperref}
\usepackage{upgreek}
\usepackage{rotating}
\usepackage{tablefootnote}
\usepackage{threeparttable}
\usepackage{lipsum}
\usepackage{graphicx}
\usepackage{ulem}
\usepackage[english]{babel}
\usepackage{color}

\let\oldequation\equation
\let\oldendequation\endequation

\renewenvironment{equation}
  {\linenomathNonumbers\oldequation}
  {\oldendequation\endlinenomath}

\begin{document}
\normalsize
\parskip=5pt plus 1pt minus 1pt

\title{\boldmath Measurement of the Born cross sections for $e^+e^- \to \eta^\prime \pi^{+}\pi^{-}$ at center-of-mass energies between $2.00$ and $3.08$~GeV}

\author{
M.~Ablikim$^{1}$, M.~N.~Achasov$^{10,c}$, P.~Adlarson$^{67}$, S. ~Ahmed$^{15}$, M.~Albrecht$^{4}$, R.~Aliberti$^{28}$, A.~Amoroso$^{66a,66c}$, Q.~An$^{63,50}$, ~Anita$^{21}$, X.~H.~Bai$^{57}$
% ...
\\
\vspace{0.2cm}
(BESIII Collaboration)\\
\vspace{0.2cm} {\it
$^{1}$Institute of High Energy Physics, Beijing 100049, People's Republic of China\\
% ...
$^{4}$Bochum Ruhr-University, D-44780 Bochum, Germany\\
% ...
$^{10}$G.I. Budker Institute of Nuclear Physics SB RAS (BINP), Novosibirsk 630090, Russia\\
% ...
$^{15}$Helmholtz Institute Mainz, Johann-Joachim-Becher-Weg 45, D-55099 Mainz, Germany\\
% ...
$^{21}$Indian Institute of Technology Madras, Chennai 600036, India\\
% ...
$^{28}$Johannes Gutenberg University of Mainz, Johann-Joachim-Becher-Weg 45, D-55099 Mainz, Germany\\
% ...
$^{50}$State Key Laboratory of Particle Detection and Electronics, Beijing 100049, Hefei 230026, People's Republic of China\\
% ...
$^{57}$University of Jinan, Jinan 250022, People's Republic of China\\
% ...
$^{63}$University of Science and Technology of China, Hefei 230026, People's Republic of China\\
% ...
$^{66a}$University of Turin, I-10125, Turin, Italy\\
$^{66b}$University of Eastern Piedmont, I-15121, Alessandria, Italy\\
$^{66c}$INFN, I-10125, Turin, Italy\\
$^{67}$Uppsala University, Box 516, SE-75120 Uppsala, Sweden\\
% ...
$^{c}$Also at the Novosibirsk State University, Novosibirsk, 630090, Russia\\
% ...
}
}

\date{\today}

\begin{abstract}

  The Born cross sections for the process $e^+e^- \to \eta^\prime \pi^{+}\pi^{-}$ at different center-of-mass energies
  between $2.00$ and $3.08$~GeV are reported with improved precision
  from an analysis of data samples
  collected with the BESIII detector operating at the BEPCII storage ring.
  An obvious structure is observed in the Born cross section line shape.
  Fit as a Breit-Wigner resonance, it has a statistical significance of
  $6.3\sigma$ and a  mass and width of $M=(2111\pm43\pm25)$~MeV/$c^2$ and
  $\Gamma=(135\pm34\pm30)$~MeV, where the uncertainties are
  statistical and systematic, respectively.
  These measured resonance parameters agree with the measurements of BABAR
  in $e^+e^- \to \eta^\prime \pi^{+}\pi^{-}$ and BESIII
  in $e^+e^- \to \omega\pi^0$ within two standard deviations.

\end{abstract}

\maketitle

\section{Introduction}

Low energy $e^+e^-$ collision experiments, where $\rho$, $\omega$ and $\phi$
resonances as well as their excited states are produced copiously, offer an ideal
test-bed to thoroughly investigate the properties of these resonances.
Many experimental results regarding these states have been summarized in the
Particle Data Group (PDG) review~\cite{PDG}.
Still, the properties of some states are still ambiguous.
Notably, the status of the $\rho(2000)$, $\rho(2150)$, and $\rho(2270)$
states is unclear, due to insufficient experimental information.
The $\rho(2000)$ was found in $p\bar{p}$ collisions~\cite{Hasan1994,Anisovich491,Anisovich508,Anisovich513,Anisovich2002},
and it was explained as a radial excitation of the $\rho(1700)$~\cite{Anisovich2002}
or a mixed state with a significant $^3D_1$ component~\cite{Bugg2004}.
The $\rho(2150)$ was initially regarded as a $2^3D_1$
state~\cite{Godfrey1985}, but later was considered to be a $4^3S_1$ state~\cite{Bugg2013,Anisovich051502,Masjuan2013,He2013,arXiv:2102.05356}.
The $\rho(2270)$ was first observed in photo-production~\cite{Atkinson1985}
and categorized as a $3^3D_1$ state~\cite{He2013}.
There are no published results
on $\rho(2000)$ and $\rho(2270)$ from $e^+e^-$ collision experiments.
The $\rho(2150)$ has been widely studied in $e^+e^-$,
$p\bar{p}$, $s$-channel $N\bar{N}$ and $\pi p$ collision experiments,
but inconsistencies in the measured masses and widths make
the $\rho(2150)$ more controversial.

According to the vector meson dominance model~\cite{VMD}, the isovector part of
the electromagnetic current in the positive G-parity process
$e^+e^-\to \eta^{\prime}\pi^+\pi^-$ allows direct production of $\rho-$like states.
Therefore this process can be used to extract the resonance parameters
of the $\rho-$like states.
The BABAR Collaboration has measured the Born cross section line shape of
$e^+ e^- \to \eta^{\prime}\pi^+\pi^-$ at center-of-mass (c.m.)
energies ($\sqrt{s}$) from 1.58 to 3.42~GeV with
the initial state radiation (ISR) technique.
A resonance-like structure around 2.1~GeV/$c^2$ reported by BABAR
could be interpreted as the $\rho(2150)$~\cite{BarBarCS}.

By using 19 data sets taken at $\sqrt{s}$ between
$2.00$ and $3.08$ GeV, the BESIII Collaboration recently reported
a $J^{\rm PC} = 1^{--}$ vector state, $Y(2040)$, in $e^+e^- \to\omega\pi^0$~\cite{Dong2020}
with a mass and width of ($2034\pm13\pm9$)~MeV/$c^2$ and ($234\pm30\pm25$)~MeV, respectively.
Here, we report the Born cross sections for $e^+e^-\to \eta^{\prime}\pi^+\pi^-$
based on the same data sets.
By fitting to these cross sections,
we measure the parameters of the possible $\rho-$like resonances.

\section{BESIII AND MONTE CARLO SIMULATION }

BESIII is a general-purpose detector located at the Beijing Electron Positron Collider (BEPCII)~\cite{BEPCII} and is designed for studies of hadron spectroscopy and $\tau$-charm physics~\cite{besphysics,besphysicsv}.
The cylindrical detector has a geometrical acceptance of 93\% of $4\pi$ solid angle and consists of four main components:

(i) A small cell, helium-based main drift chamber~(MDC) with 43 layers immersed in the 1.0~T magnetic field of a super conducting solenoid. The average single-hit resolution is
135 $\upmu$m, and the charged particle momentum resolution is 0.5\% at 1.0~GeV/$c$.

(ii) A time-of-flight system (TOF) made from two layers of plastic scintillator, with 88 counters 5~cm thick and 2.4~m long in each layer for the barrel, and 96 fan-shaped counters in each
end cap. It provides timing information with a resolution of 68~ps in the barrel and 110~ps
in the end caps, which yields 2$\sigma$ $K/\pi$ separation at 1.0~GeV/$c$.

(iii) An electromagnetic calorimeter (EMC) consisting of 6240 CsI (Tl) crystals
in a cylindrical barrel and two end caps to measure shower energies.
The photon energy resolution at 1.0~GeV is 2.5\% in the barrel
and 5\% in the end caps, while the position
resolution is 6 mm and 9 mm for barrel and end caps, respectively.

(iv) A resistive plate chamber (RPC)-based muon chamber (MUC) with nine layers in
the barrel and eight layers in the end caps providing 2 cm position resolution.

Monte Carlo (MC) simulations of the full detector, based on {\sc geant4}~\cite{Geant4}
simulation software, and the BESIII {\sc Object Oriented Simulation Tool}
(BOOST)~\cite{Boost}, are used
to optimize the event selection criteria, understand potential backgrounds,
and determine the detection efficiency. The BOOST package contains the detector geometry
and material description, the detector response and signal digitization models,
as well as records of the detector running conditions and performance.
Large inclusive MC samples at $\sqrt{s}=2.1250$ and 2.3960~GeV are generated to estimate potential backgrounds.
The processes  $e^+e^-\to e^+e^-, \mu^+\mu^-$ and $\gamma\gamma$ are generated with the Babayaga
generator~\cite{Babayaga}, while $e^+e^- \to$ hadrons and two photon processes are simulated by
the {\sc luarlw}~\cite{LUARLW} and {\sc bestwogam}~\cite{BESTWOGAM} generators, respectively.
Due to the dominance of $\rho\eta'$ in $e^+e^- \to \eta^\prime\pi^{+}\pi^{-}$
process~\cite{BarBarCS}, the signal MC is
generated with {\sc conexc}~\cite{ConExc} as $e^+e^- \to \rho\eta^\prime$.
The cross sections measured by the BABAR experiment~\cite{BarBarCS}
are used as initial input.
The ISR, vacuum polarization and the angular distributions of the final state have been taken
into account in the generator.
The wide $\rho$ resonance in the intermediate state is described by the Gounaris-Sakurai (GS) model~\cite{GSi}.
The $\eta^\prime \to \gamma\pi^{+}\pi^{-}$ decay is simulated with a model
based on the results of the amplitude analysis in Ref.~\cite{Gammapipi},
and the $\eta^\prime \to\eta\pi^{+}\pi^{-}$ decay is produced
using a phase-space (PHSP) model.

\section{ Data Analysis}

In this analysis, the $\eta^\prime$ is reconstructed via the two decay modes
$\eta^\prime \to \eta\pi^{+}\pi^{-}$ and $\eta^\prime \to\gamma\pi^{+}\pi^{-}$, which will be referred to as mode I and mode II, respectively.
For mode I, the $\eta$ is reconstructed in the decay $\eta \to \gamma\gamma$.

Charged tracks are reconstructed using hits in the MDC.
Each track is required to be within the polar angle ($\theta$)
region $|\cos\theta|<0.93$ and have a distance of  closest approach to
the interaction point within $\pm$10~cm along the beam direction (z-axis)
and within 1~cm in the transverse plane. For both modes, it is required that
there are exactly four charged tracks with net zero charge.
Particle identification (PID) for charged tracks combines the measurements
of the specific ionization energy, $dE/dx$, in the MDC and the flight time in the TOF.
The charged track is identified as a pion if the confidence level for the pion hypothesis is
greater than those for both the kaon and proton hypotheses.

Showers in the EMC are chosen as photon candidates
if they satisfy the following requirements:
the deposited energy must be larger than 25~MeV in the barrel region
($|\cos\theta|<0.80$) and 50~MeV in the end cap ($0.86<|\cos\theta|<0.92$).
To suppress electronic noise and showers unrelated
to the event, the EMC time deviation from the event start time is required
to be within (0,700)~ns.
For mode I (mode II), it is required that there are at least two (one)
photons.

\begin{figure}[htbp]
\begin{center}
\begin{overpic}[width=7.5cm]{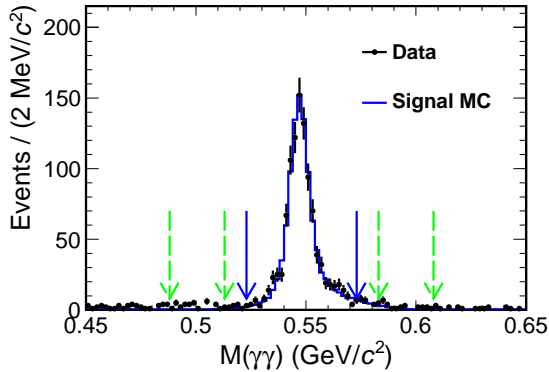}
\end{overpic}
\end{center}
\vspace*{-0.6cm}
\caption{ The $M(\gamma\gamma)$ spectrum
of the $e^+e^-\to \eta^\prime \pi^+\pi^-$ candidate events for mode I
in data (dots with error bars) and signal MC (histogram) at $\sqrt{s}=$2.1250~GeV.
The region between the two solid blue arrows is the $\eta$ signal region,
while the regions between the pairs of dashed green arrows are the $\eta$ sideband regions.
}
\label{etamass}
\end{figure}

A vertex fit is imposed on the selected charged tracks to ensure
that they originate from the same interaction point.
To improve momentum resolution and to suppress background,
a four-constraint (4C) kinematic fit imposing energy-momentum conservation under the hypotheses of $e^+e^-\to \gamma\gamma\pi^{+}\pi^{-}\pi^{+}\pi^{-}$ and
$e^+e^-\to \gamma\pi^{+}\pi^{-}\pi^{+}\pi^{-}$ is employed on the selected candidates for mode~I and mode~II, respectively.
For events with additional photon(s), the combination with the smallest $\chi^2_{\rm{4C}}$ is retained.
Based on an optimization of $s/\sqrt{s'+b}$ for the requirement on $\chi^2_{\rm{4C}}$,
where $s$ and $s'+b$ is the number of events in the $\eta'$ signal region
(($M(\eta\pi^+\pi^-, \gamma\pi^+\pi^-)\in(0.943,0.973)$~GeV/$c^2$)) in signal MC and data, respectively,
candidate events with $\chi^2_{\rm{4C}}<100$ ($\chi^2_{\rm {4C}}< 50$)
for mode I (mode II) are accepted for further analysis.
For mode II, to suppress contaminations from $e^{+}e^{-} \to 2(\pi^+\pi^-)$
and $e^{+}e^{-} \to 2(\gamma\pi^{+}\pi^{-})$,
two additional 4C kinematic fits under each of these two hypotheses
are performed.
Events are discarded if the $\chi^2_{\rm {4C}}$ for either of these
fits is less than the signal mode 4C kinematic fit.
To further suppress background from $e^{+}e^{-} \to 2(\pi^+\pi^-)$,
all photon energies are required to be greater than 0.1~GeV for mode II.
For convenience, we take the data set at $\sqrt s=$2.1250~GeV,
which has the the largest statistics, as an example in this section.
Figure~\ref{etamass} shows
the $M(\gamma\gamma)$ distribution of the accepted candidates
for $e^+e^-\to \eta^\prime \pi^+\pi^-$ using mode I.
The events in the $\eta$ mass signal region,
$M(\gamma\gamma)\in (0.523,0.573)$~GeV/$c^2$
(the region between two solid blue arrows), are kept for further analysis.

\begin{figure}[htbp]
\begin{center}
  \begin{overpic}[width=7.5cm]{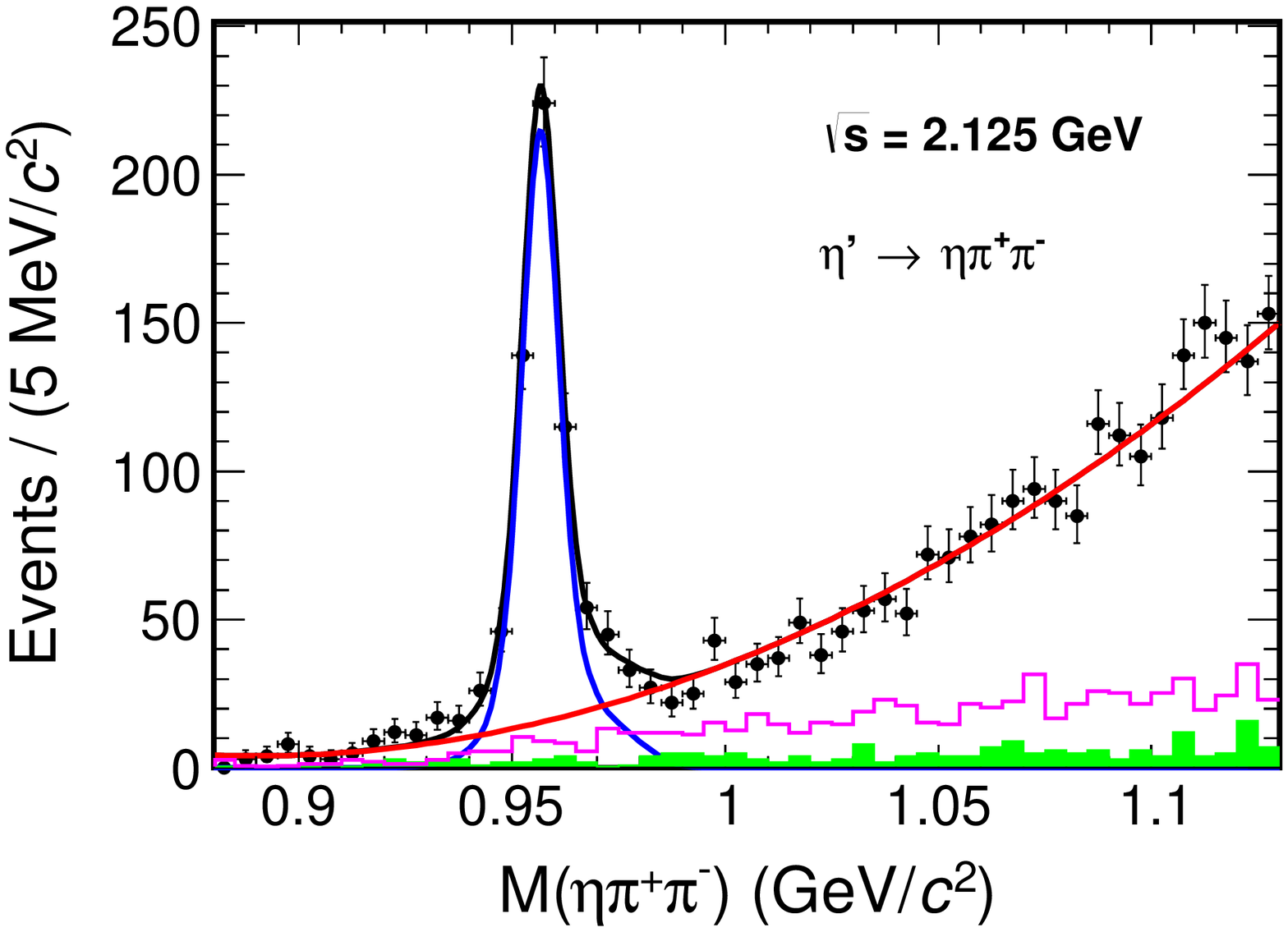}

\put(25,55){(a)}
\end{overpic}
    \begin{overpic}[width=7.5cm]{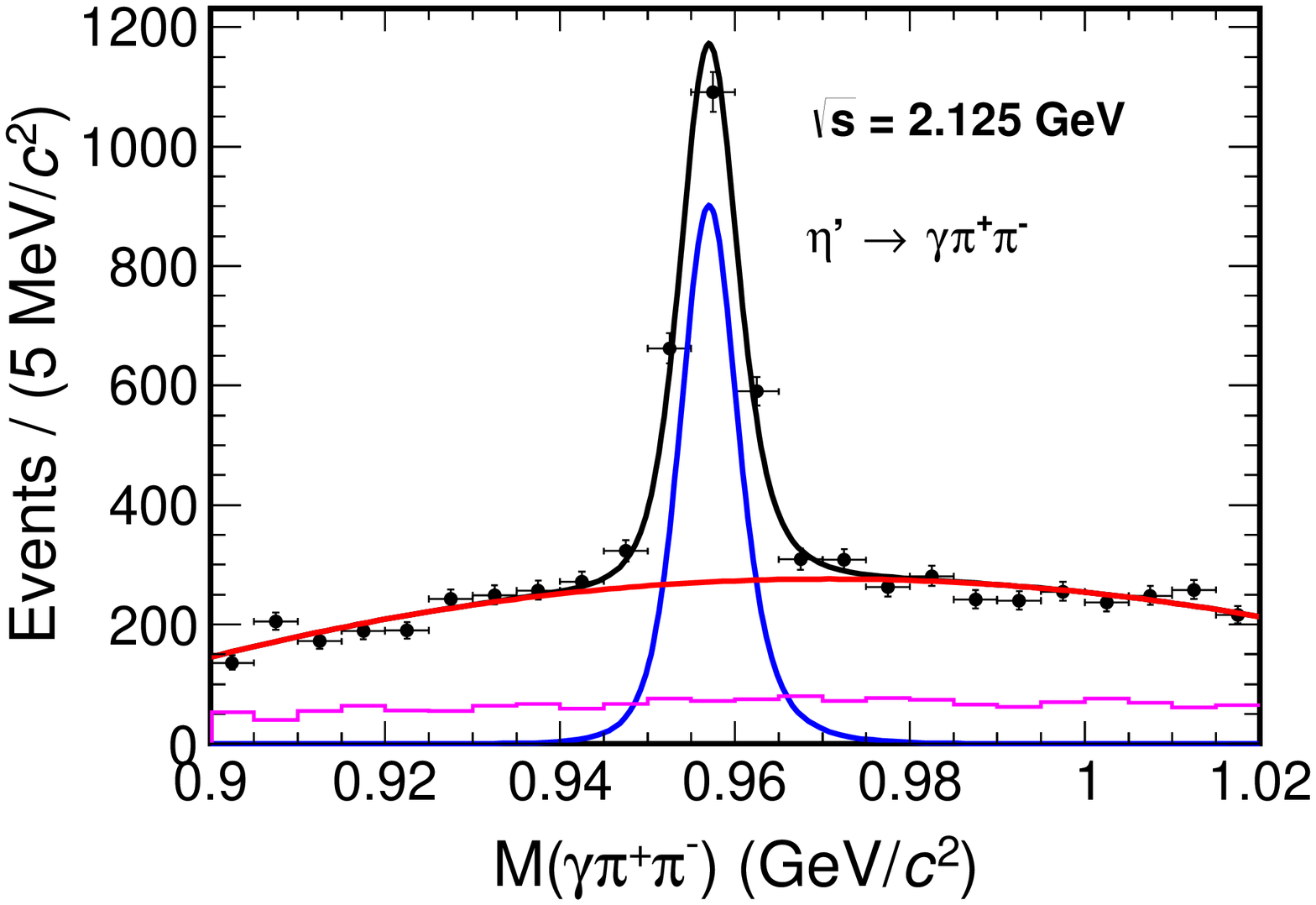}

\put(25,55){(b)}
\end{overpic}
\end{center}
\vspace*{-0.5cm}
\caption{Fits to (a) $M(\eta\pi^{+}\pi^{-})$ and (b) $M(\gamma\pi^{+}\pi^{-})$  for
the accepted candidate events in data at $\sqrt{s}=$2.1250~GeV.
Dots with error bars show the data, black lines give the total fit results,
blue dotted lines are the signal components, red lines are the smooth backgrounds, and
magenta histograms are the inclusive MC samples.
The inclusive MC samples are normalized to the data luminosity.
The green shaded histogram in (a) shows the events from the $\eta$ sidebands in data.
 }
\label{etapmass}
\end{figure}

Figure~\ref{etapmass} presents the distributions of $M(\eta\pi^+\pi^-)$ for mode I and $M(\gamma\pi^+\pi^-)$ for mode II of the accepted candidates.
Clear $\eta^\prime$ signals are observed.
For mode I, the non-$\eta$ backgrounds in the $\eta$ mass signal region are examined by the events in the $\eta$ mass sideband region,
which is defined as $M(\gamma\gamma)\in(0.488,0.513)\cup (0.583,0.608)$~GeV/$c^2$ (the region between two neighboring dashed green arrows in Figure.~\ref{etamass}). The resulting $M(\eta \pi^+\pi^-)$ distribution of the $\eta$ sideband events is shown as the green shaded histogram in Figure.~\ref{etapmass}\,(a).
Further studies based on the inclusive MC samples show that
the dominant backgrounds are
$e^+e^- \to \eta 2(\pi^+ \pi^- )|_{\rm non{\text-}\eta^\prime}$ for mode I and
$e^{+}e^{-} \to 2(\pi^+\pi^-)$ for mode II.
The resulting $M(\eta\pi^+\pi^-, \gamma\pi^+\pi^-)$ distribution of the
accepted background events from the inclusive MC samples for mode I and
mode II are shown as the magenta histograms of Figure.~\ref{etapmass}\,
(a) and (b), respectively. No peaking backgrounds are seen near the $\eta^\prime$ mass.
Therefore, we will fit the $M(\eta\pi^+\pi^-)$ ($M(\gamma\pi^+\pi^-)$) spectra
using a smooth background shape to account for the remaining backgrounds in the analysis.

\begin{figure}[htbp]
\begin{center}
\begin{overpic}[width=7.5cm]{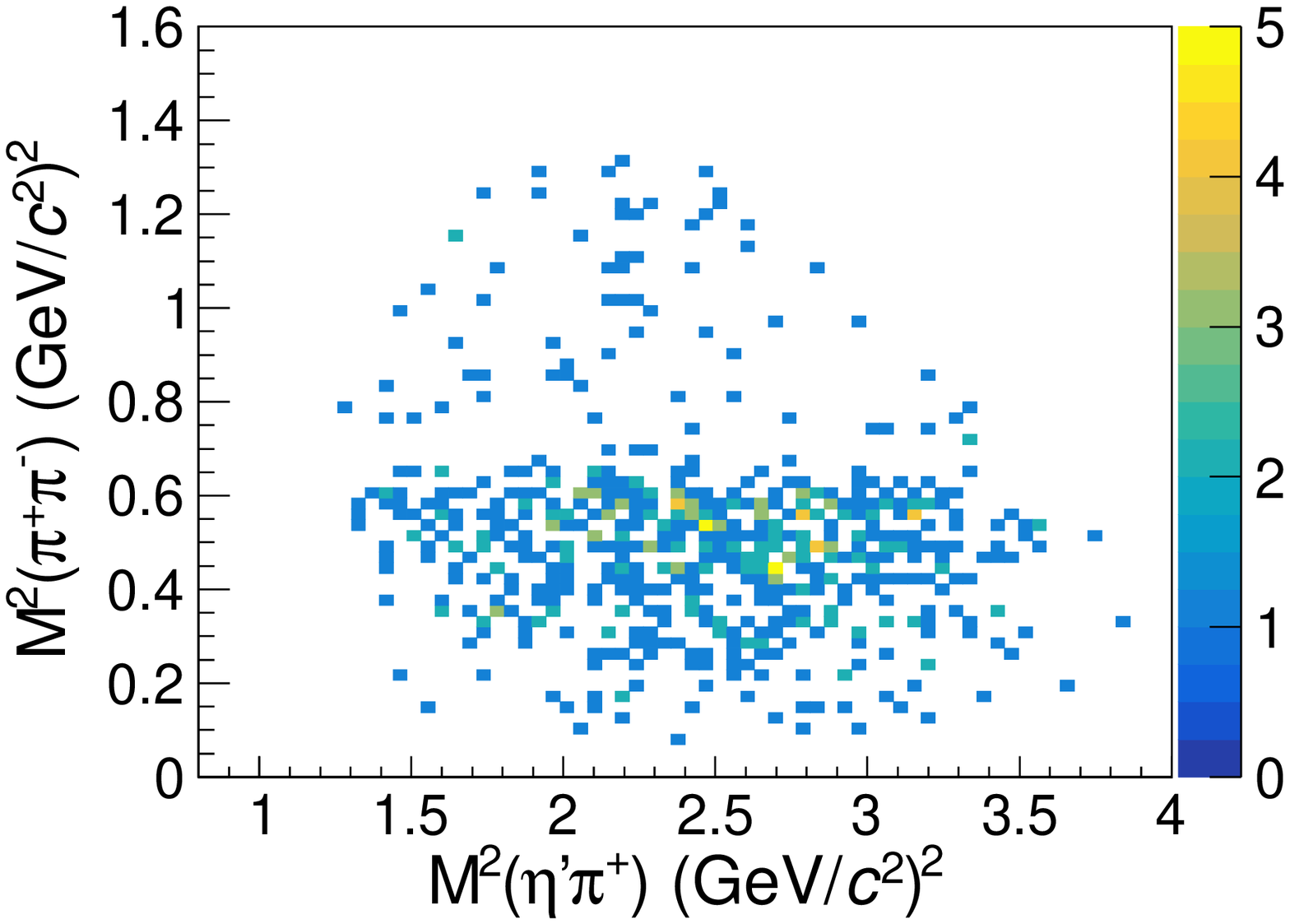}
 \put(70,57){(a)}
\end{overpic}
\begin{overpic}[width=7.5cm]{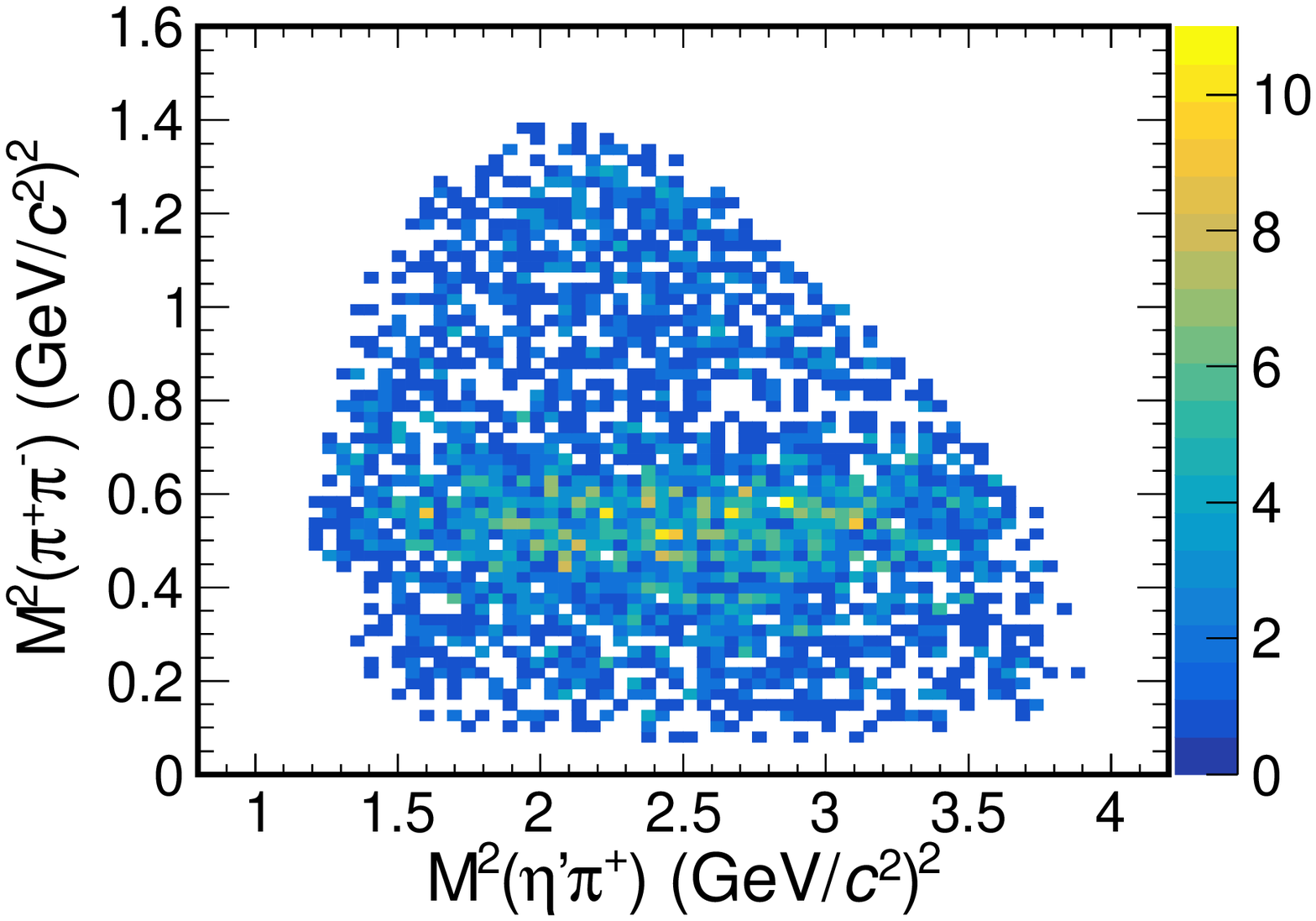}
 \put(70,57){(b)}
\end{overpic}
\end{center}
\vspace*{-0.6cm}
\caption{Dalitz plots of the accepted $e^+e^-\to \eta^\prime \pi^+\pi^-$ candidate events
for (a) mode I and (b) mode II in data at $\sqrt{s}=$2.1250~GeV.
The requirement $M(\eta\pi^+\pi^-, \gamma\pi^+\pi^-)\in(0.943,0.973)$~GeV/$c^2$ has been imposed.}
\label{Dalitzplot}
\end{figure}

\begin{figure}[htbp]
\begin{center}
\begin{overpic}[width=7.5cm]{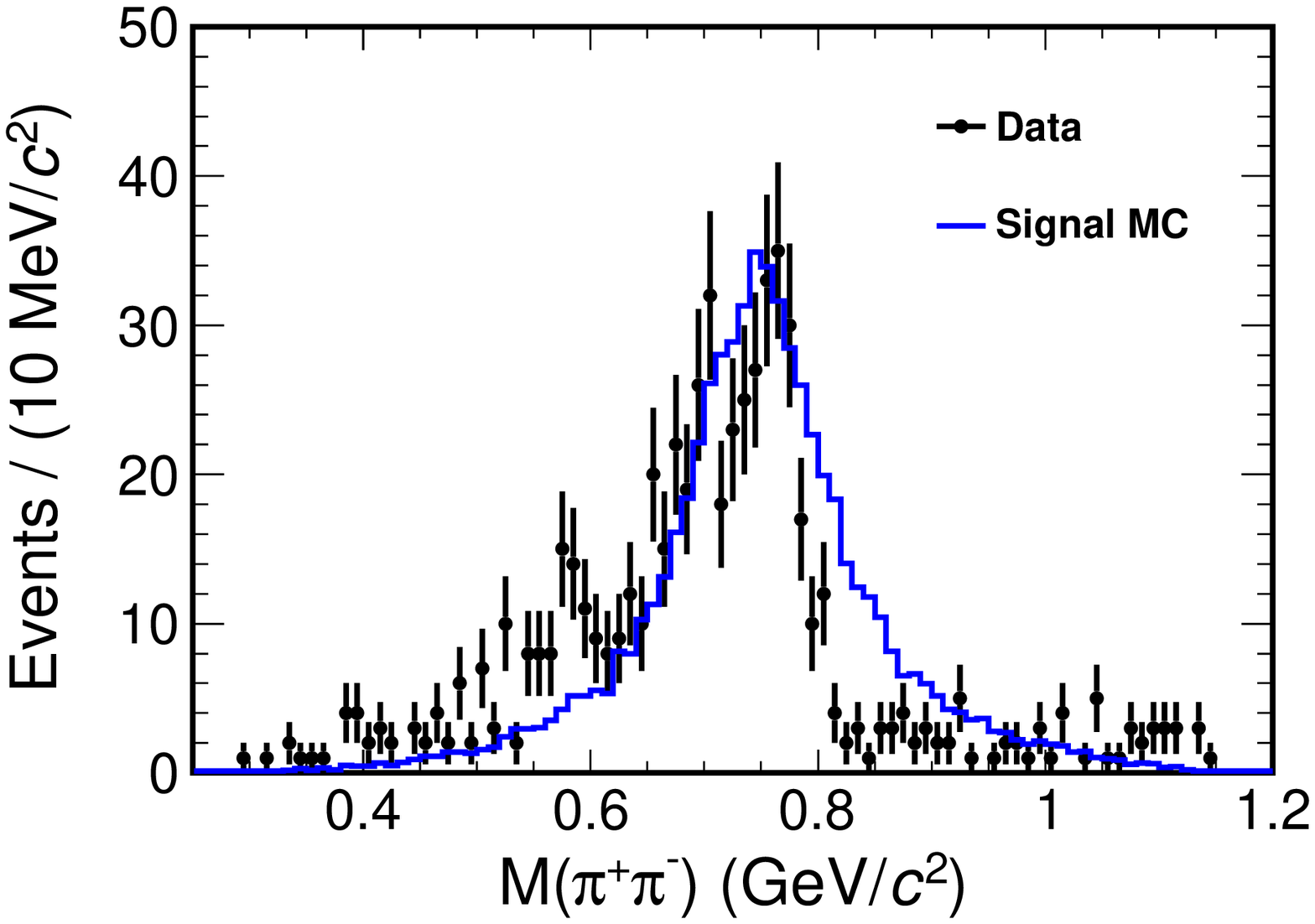}
 \put(25,55){(a)}
\end{overpic}
\begin{overpic}[width=7.5cm]{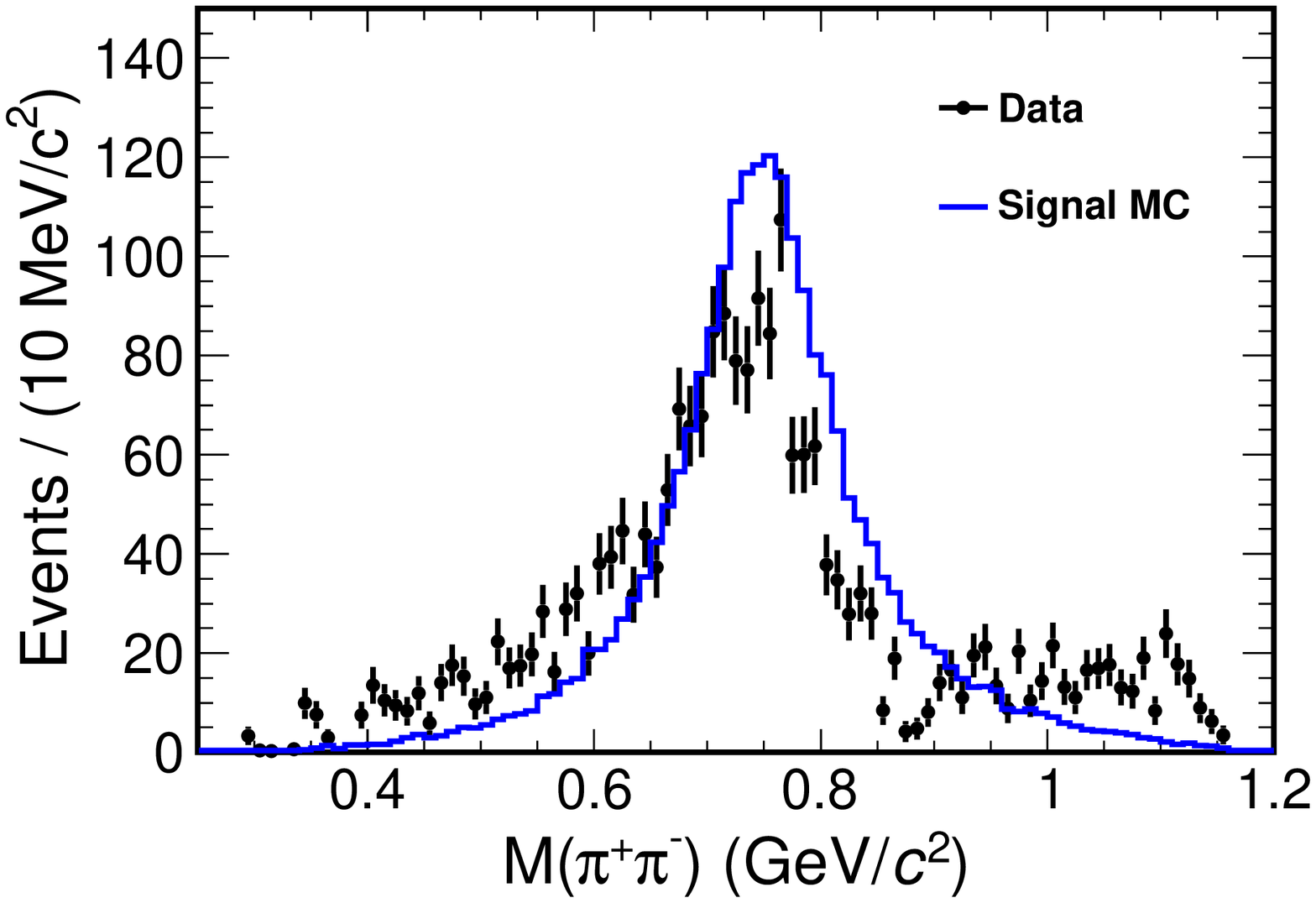}
 \put(25,55){(b)}
\end{overpic}
\end{center}
\vspace*{-0.6cm}
\caption{Projections of the corresponding Dalitz plots on $M^2(\pi^+\pi^-)$ axis
for (a) mode I and (b) mode II in data (dots with error bars) and signal MC (histograms) at $\sqrt{s}=$2.1250~GeV.
}
\label{Mpipi}
\end{figure}

Possible intermediate states in $e^+e^- \to \eta^\prime \pi^{+}\pi^{-}$ are examined by
the corresponding Dalitz plots selecting the $\eta^\prime$ signal region as
$M(\eta\pi^+\pi^-, \gamma\pi^+\pi^-)\in(0.943,0.973)$~GeV/$c^2$, shown in Figure~\ref{Dalitzplot}.
After subtracting the non-$\eta'$ background in the $\eta'$ sideband region,
$M(\eta\pi^+\pi^-, \gamma\pi^+\pi^-)\in(0.918,0.938)\cup (0.978,0.998)$~GeV/$c^2$,
with a weight factor of 0.8.
Figure~\ref{Mpipi} presents the projections of the corresponding Dalitz plots on the
$M^2(\pi^+\pi^-)$ axis for two modes.
As expected, the dominant component is $e^+e^-\to \rho\eta^\prime$,
and the non-$\rho$ contribution is less than 10\%.
A fit to the $M(\pi^+\pi^-)$ in the high-statistics bin at c.m. energy of
2.125~GeV shows that the apparent shift between MC and data of the $\rho(770)$
peak can be explained by interference between $\rho(770)$ and non-$\rho(770)$
process.

\section{Born Cross section Measurement}

The Born cross section, $\sigma^{\rm B}$, at each c.m.~energy is determined as
\begin{equation}
\label{eq:sigmaobs}
\sigma^{\rm B} = \frac{N^{\rm obs}}{\mathcal{L}\cdot\epsilon \cdot {\mathcal B}\cdot (1+\delta^{\mathit{r}})\cdot \frac{1}{|1-\Pi|^2}} ,
\end{equation}
where $N^{\rm{obs}}$ is the signal yield,
$\mathcal{L}$ is the integrated luminosity of the data set,
$\epsilon$ is the detection efficiency, ${\mathcal B}$ is the product
of the relevant daughter branching fractions, i.e.,
${\mathcal B}={\mathcal B}(\eta'\to\eta\pi^{+}\pi^{-})\cdot{\mathcal B}(\eta\to\gamma\gamma)=16.8\%$
for mode I and ${\mathcal B}={\mathcal B}(\eta'\to\gamma\pi^{+}\pi^{-})=28.9\%$ for mode II~\cite{PDG}.
The factor $(1+\delta^{\mathit{r}})$ is the ISR correction factor and
$\frac{1}{|1-\Pi|^2}$ is the vacuum polarization factor.
Both $(1+\delta^{\mathit{r}})$ and $\frac{1}{|1-\Pi|^2}$
are obtained from MC simulations~\cite{ISR,VP}.
To obtain reliable detection efficiencies and ISR factors, the
Born cross sections used as input in the generator have been iterated until the product
$(1+\delta^{\mathit{r}})\cdot\epsilon$ has converged, defined as an iteration leading to a relative change of less than 1.0\%.

The signal yields are obtained from a simultaneous unbinned maximum-likelihood fit
to the $M(\eta\pi^+\pi^-)$ and $M(\gamma\pi^+\pi^-)$ spectra at each c.m.~energy.
The signal is described by a MC-simulated shape convolved with a
Gaussian function.
The parameters of the Gaussian function are free.
Among the different data sets, a common Gaussian convolution is used to
compensate for potential differences in calibration and resolution between data and MC simulation.
A second-order Chebychev polynomial is used to describe the combinatorial background shape.
In the fit, the two modes share the same Born cross section,
and the expected signal yields are $N^{obs}=\sigma^B\cdot\mathcal{L}\cdot\epsilon \cdot {\mathcal B}\cdot (1+\delta^{\mathit{r}})\cdot \frac{1}{|1-\Pi|^2}$.
Figure~\ref{etapmass} shows the fit result for data taken at $\sqrt{s}=2.1250$~GeV.
Similar combined fits to the two final states are performed for each c.m.~energy;
the resulting cross sections and related variables are listed
in Table~\ref{CrossSectionTab}.
They are consistent with those from the BABAR experiment~\cite{BarBarCS},
as seen in Figure.~\ref{fig:lineshapefit}.
Independent fits to the $M(\eta\pi^+\pi^-)$ and $M(\gamma\pi^+\pi^-)$ spectra
are also performed and the individual signal yields are also summarized
in Table~\ref{CrossSectionTab}.

\begin{table*}[htbp].
\begin{center}
 \caption{Summary of the integrated luminosities ($\mathcal{L}$)~\cite{LuminosityFinal}, observed event yields from independent fit ($N^{obs}$), detection efficiencies ($\epsilon$), radiative correction factors $(1+\delta^{\gamma})$, vacuum polarization factors $\frac{1}{|1-\Pi|^2}$, and the obtained Born cross section ($\sigma^{\rm B}$) at different c.m.~energies ($\sqrt{s}$). The first uncertainties for $\sigma^{\rm B}$ are statistical and the second are systematic; those for $N^{\rm obs~I}$ and $N^{\rm obs~II}$ are statistical only. The symbols of I and II represent the $e^+e^- \to \eta^\prime\pi^+\pi^-$ processes reconstructed via $\eta^\prime \to \eta\pi^+\pi^-$ and $\eta^\prime \to \gamma\pi^+\pi^-$, respectively.}
\label{CrossSectionTab}
\begin{tabular}{ccccccccc}  \hline
$\sqrt{s}$~(GeV)  & $\mathcal{L}$~(pb$^{-1}$) & $N^{\rm obs~I}$  & $N^{\rm obs~II}$  & $\epsilon^{\rm I}$  & $\epsilon^{\rm II}$  & $1+\delta^{\mathit{r}}$ &$\frac{1}{|1-\Pi|^2}$   & $\sigma^{\rm B}$ (pb)   \\ \hline
2.0000   & 10.1  &$35.4 \pm 7.5$  &$119.2 \pm 16.8$   &0.157  &0.265   & 0.983     & 1.037      & $144.0  \pm  17.0  \pm 9.7 $ \\
2.0500	 & 3.34   &$23.9 \pm 5.5$  &$55.6 \pm 10.6$    & 0.167	&0.275   & 0.941	 & 1.038	  & $229.3  \pm  33.2  \pm 15.7 $ \\
2.1000	 & 12.2  &$58.6 \pm 9.1$  &$207.9 \pm 19.6$   & 0.166	&0.269   & 0.979	 & 1.039	  & $200.1  \pm  16.3  \pm 14.1 $ \\
2.1250	 & 108.  &$555.4\pm27.4$	&$1684.6\pm58.2$    & 0.160	&0.258   & 1.016	 & 1.039	  & $191.0  \pm  5.4  \pm 12.7 $ \\
2.1500	 & 2.84   &$8.7  \pm3.7$	&$53.0 \pm 9.3$     & 0.154	&0.252   & 1.045	 & 1.040	  & $184.8  \pm  31.9  \pm 12.5 $ \\
2.1750	 & 10.6  &$43.7 \pm 7.7$  &$116.4 \pm 15.7$   & 0.159	&0.251   & 1.059     & 1.040	  & $137.6  \pm  14.7  \pm 9.1 $ \\
2.2000	 & 13.7  &$36.0 \pm 7.6$	&$130.1 \pm 16.6$   & 0.155	&0.247   & 1.075	 & 1.040	  & $108.8  \pm  12.0  \pm 9.3 $ \\
2.2324	 & 11.9  &$35.1 \pm 7.4$	&$127.0 \pm 15.8$   & 0.156	&0.244   & 1.089	 & 1.041	  & $122.0  \pm  13.2  \pm 9.9 $ \\
2.3094	 & 21.1  &$45.8 \pm 8.2$	&$149.1 \pm 17.9$   & 0.153	&0.237   & 1.109	 & 1.041	  & $83.3  \pm  8.4  \pm5.6 $  \\
2.3864	 & 22.5  &$57.5 \pm 8.8$	&$158.1 \pm 17.4$   & 0.152	&0.232   & 1.120	 & 1.041	  & $87.9  \pm  7.9  \pm6.3 $  \\
2.3960	 & 66.9  &$167.9\pm15.2$	&$496.1 \pm 31.7$   & 0.156	&0.235   & 1.119	 & 1.041	  & $89.3  \pm  4.7  \pm5.9 $ \\
2.6444	 & 33.7  &$46.8 \pm 7.9$  &$164.1 \pm 16.7$   & 0.160	&0.223   & 1.154	 & 1.039	  & $55.4  \pm  4.9  \pm3.6 $ \\
2.6464	 & 34.0  &$51.8 \pm 8.3$	&$148.2 \pm 17.0$   & 0.160	&0.222   & 1.157	 & 1.039	  & $52.7  \pm  5.0  \pm3.8 $  \\
2.9000	 & 105.  &$137.4\pm13.0$	&$288.7 \pm 22.4$   & 0.163	&0.211   & 1.192	 & 1.033	  & $37.3  \pm  2.2  \pm2.8 $  \\
2.9500	 & 15.9  &$19.6 \pm 4.9$	&$43.2 \pm 8.3$     & 0.161	&0.203   & 1.197	 & 1.029	  & $37.4  \pm  5.7  \pm2.9 $  \\
2.9810	 & 16.1   &$13.5 \pm 4.3$	&$35.5 \pm 8.3$     & 0.163	&0.206   & 1.199	 & 1.025	  & $28.1  \pm  5.2  \pm2.0 $  \\
3.0000	 & 15.9   &$16.1 \pm 4.5$	&$33.8 \pm 7.8$     & 0.163	&0.204   & 1.197	 & 1.021	  & $29.7  \pm  5.3  \pm2.0 $ \\
3.0200	 & 17.3   &$13.6 \pm 4.8$	&$33.9 \pm 7.6$     & 0.163 &0.204   & 1.199     & 1.014      & $26.1  \pm  5.0  \pm2.3 $  \\
3.0800	 & 126.  &$88.0 \pm10.2$  &$218.3\pm 20.1$    & 0.157	&0.191   & 1.135	 & 0.915	  & $28.0  \pm  2.0  \pm1.9 $ \\
\hline
 \end{tabular}
\end{center}
\end{table*}

By implementing the same strategy described
in~\cite{prd0320012019,Yankun,Yateng,Yankun2020}, several sources of
systematic uncertainty on the measured cross section are considered.
These uncertainties for all datasets are summarized in Table~\ref{tab:sysuncer}.

The uncertainty in the measurement of the integrated luminosity of the data set has been determined to be $1.0\%$~\cite{LuminosityFinal}.
The uncertainties of the track reconstruction and PID efficiencies of charged pions
have been studied by using a control sample of
$e^+e^- \to K^+K^-\pi^+\pi^-$ \cite{prd0320012019},
resulting in $1.0$\% per pion for tracking and $1.0$\% per pion for PID.
The uncertainty on the photon efficiency is estimated as $1.0$\% per
photon by using a control sample of $e^+e^- \to \pi^+\pi^-\pi^0$~\cite{photonerror}.
Since the number of photon for mode I and mode II are two and one respectively,
the larger change in the cross section due to shifting the detection efficiency
by $\pm2.0$\% for mode~I and $\pm1.0$\% for mode~II in the simultaneous fit
(with both shifts having the same sign),  is taken as the systematic uncertainty.

To estimate the uncertainty of the MC modeling, we examine the efficiency
at 2.1250 GeV,
which has the largest statistics among all the c.m.~energies. The signal MC
samples are weighted according to the
the Dalitz distribution of $e^+e^- \to \eta^\prime\pi^{+}\pi^{-}$ in data after background subtraction.
The difference between the weighted efficiency and the nominal efficiency,
1.3\%, is taken as the systematic uncertainty.
Due to the limited statistics at other c.m.~energies, this systematic uncertainty is taken to be the same as that for 2.1250 GeV at all energies.

The track helix parameters for the MC simulation are corrected before the 4C kinematic
fit to account for observed data-MC differences~\cite{Aixiaocong}.
The differences in detection efficiencies with and without corrections,
($0.8-1.9$)\%,are assigned as the systematic uncertainty
from the 4C kinematic fit.

The uncertainty originating from the ISR correction factor is taken as
the relative difference of the values of $(1+\delta^{\mathit{r}})\cdot\epsilon$
between the last two iterations of the cross section measurement.

Three uncertainties associated with the fits to the mass spectra are examined.
The background shape is replaced by a third-order Chebychev polynomial function.
The signal shape is replaced with a Gaussian function with fixed resolution,
obtained by shifting the resolution from the nominal fit by one standard deviation in each direction.
Finally, the fit range is varied by $\pm5$~MeV.
The quadrature sum of the changes in the fitted yields is taken as the uncertainty.

The impact of uncertainties on the branching fractions of the intermediate
states is examined by changing the branching fractions of $\eta$ and $\eta^\prime$ by $\pm1\sigma$ in the simultaneous fit,
where $\sigma$ is the uncertainty of the individual branching fractions.
The difference on the cross section, $1.8$\%, is taken as the systematic uncertainty.

Adding the systematic uncertainties in quadrature yields the total systematic uncertainties of
the measured Born cross sections, which are summarized in Table~\ref{tab:sysuncer}.

\begin{table*}[htbp]
\begin{center}
  \caption{Systematic uncertainties (in \%) from luminosity ($\mathcal{L}$),
photon reconstruction (photon), tracking, PID, MC modeling (MC), kinematic fit (KF),
radiation correction (RC), fitting and quoted branching fraction in the cross section
measurements. The sources with $^*$ superscript are common systematic uncertainties for
different c.m.~energies.}
  \begin{tabular}{ccccccccccccc}
  \hline

  $\sqrt{s}$~(GeV)  &$\mathcal{L}^*$  & photon$^*$  & tracking$^*$ &PID$^*$ &MC$^*$ &KF  &RC  &Fitting  &${\mathcal B}^*$ & Sum\\ \hline
2.0000    &1.0    &1.3  &4.0   &4.0  &1.3  &1.6   &0.5  &1.7  &1.8  &6.7  \\
2.0500    &1.0    &1.3  &4.0   &4.0  &1.3  &1.5   &0.1  &2.3  &1.8  &6.9  \\
2.1000    &1.0    &1.3  &4.0   &4.0  &1.3  &1.9   &0.5  &2.5  &1.8  &7.1  \\
2.1250    &1.0    &1.3  &4.0   &4.0  &1.3  &1.8   &0.4  &1.2  &1.8  &6.7  \\
2.1500    &1.0    &1.3  &4.0   &4.0  &1.3  &1.7   &0.4  &1.7  &1.8  &6.8  \\
2.1750    &1.0    &1.3  &4.0   &4.0  &1.3  &1.7   &0.1  &1.3  &1.8  &6.7  \\
2.2000    &1.0    &1.3  &4.0   &4.0  &1.3  &1.8   &0.8  &5.4  &1.8  &8.5  \\
2.2324    &1.0    &1.3  &4.0   &4.0  &1.3  &1.7   &0.1  &4.9  &1.8  &8.2  \\
2.3094    &1.0    &1.3  &4.0   &4.0  &1.3  &1.7   &0.2  &1.8  &1.8  &6.8  \\
2.3864    &1.0    &1.3  &4.0   &4.0  &1.3  &1.5   &0.6  &3.2  &1.8  &7.2  \\
2.3960    &1.0    &1.3  &4.0   &4.0  &1.3  &1.4   &0.3  &1.3  &1.8  &6.6  \\
2.6444    &1.0    &1.3  &4.0   &4.0  &1.3  &1.4   &0.5  &0.7  &1.8  &6.5  \\
2.6464    &1.0    &1.3  &4.0   &4.0  &1.3  &1.3   &0.4  &3.3  &1.8  &7.2  \\
2.9000    &1.0    &1.3  &4.0   &4.0  &1.3  &1.3   &0.1  &3.9  &1.8  &7.5  \\
2.9500    &1.0    &1.3  &4.0   &4.0  &1.3  &0.8   &0.1  &4.2  &1.8  &7.6  \\
2.9810    &1.0    &1.3  &4.0   &4.0  &1.3  &1.9   &0.1  &3.0  &1.8  &7.2  \\
3.0000    &1.0    &1.3  &4.0   &4.0  &1.3  &1.3   &0.1  &1.9  &1.8  &6.7  \\
3.0200    &1.0    &1.3  &4.0   &4.0  &1.3  &1.2   &0.4  &5.8  &1.8  &8.7  \\
3.0800    &1.0    &1.3  &4.0   &4.0  &1.3  &1.4   &0.7  &2.3  &1.8  &6.9  \\
    \hline
    \end{tabular}
    \label{tab:sysuncer}
\end{center}
\end{table*}

\section{FIT TO THE BORN CROSS SECTION}

\begin{figure}[htbp]
\begin{center}
  \begin{overpic}[width=7.5cm]{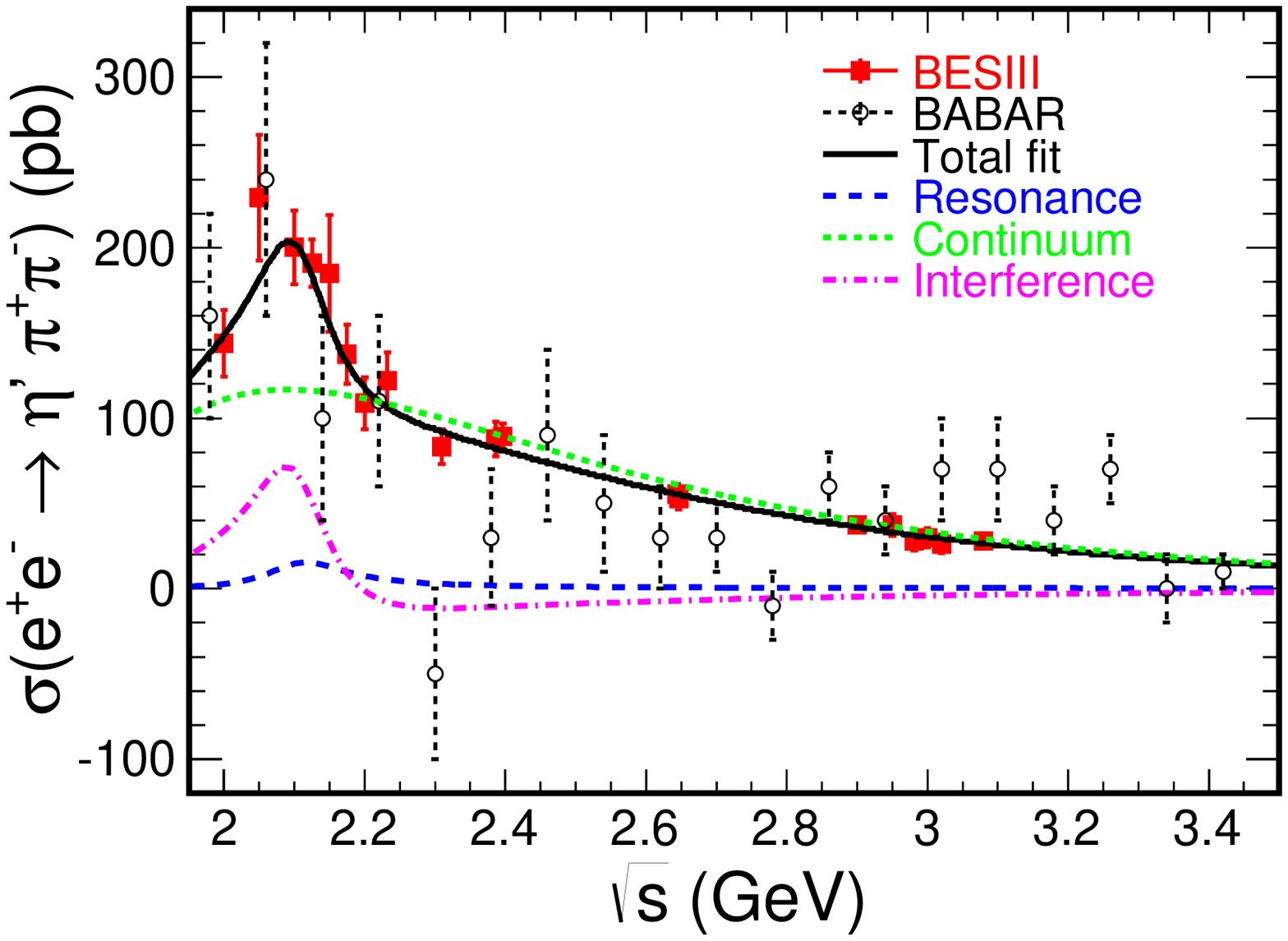}
  \put(35,63){(a)}
  \end{overpic}
    \begin{overpic}[width=7.5cm]{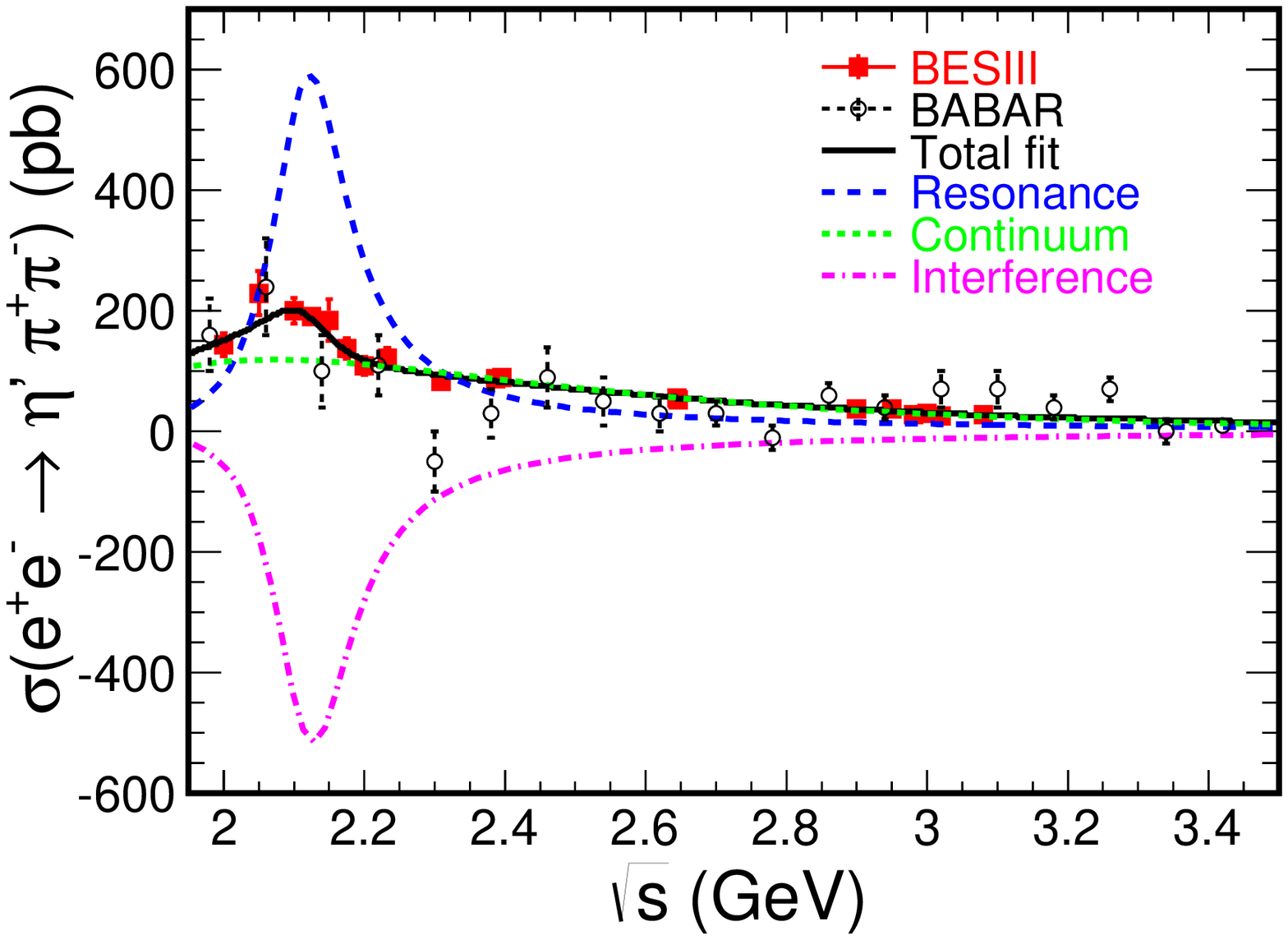}
    \put(35,63){(b)}
    \end{overpic}
\end{center}
\vspace*{-0.6cm}
\caption{Fit to the $e^+e^- \to \eta^\prime\pi^+\pi^-$ Born cross sections (only BESIII results). (a) Solution I, constructive interference. (b) Solution II, destructive interference.
Red solid dots with error bars are BESIII data, and hollow dots with error bars
are BABAR data. The black solid curve is the total fit result, the blue dashed line
is the resonant component, the green dashed line is the continuum contribution,
and the magenta dot-dashed line represents the interference between the resonance and the continuum contribution.
The systematic uncertainties are included.}
\label{fig:lineshapefit}
\end{figure}

The obtained Born cross sections are shown in Figure.~\ref{fig:lineshapefit} in which a clear structure around 2.05~GeV is observed.
To determine the mass and width of the possible resonance, a $\chi^2$ fit is performed to these cross sections.
The cross section is parameterized as the coherent sum of a resonant amplitude described by a Breit-Wigner function and an $s$-dependent continuum amplitude~\cite{s_depend}:
\begin{equation} \label{eq:fitfuntion}
\sigma(s) = \left|\frac{C_0}{s^{n}}\sqrt{\Phi(\sqrt{s})} + C_1\cdot BW(\sqrt{s}) \times e^{i\phi}\right|^{2},
\end{equation}
where $C_0$ and $n$ are the continuum parameters,
$C_1=3.894\times10^{5}\; \textrm{nb}\cdot\text{GeV}^2$ is a unit conversion factor,
and $\phi$ is the phase angle between the amplitudes.
The relativistic Breit-Wigner amplitude is given by
\begin{equation} \label{eq:BWphspfactor}
BW(\sqrt{s}) = \frac{\sqrt{12\pi\Gamma^{ee}_{R}\mathcal{B}_R\Gamma^{\rm tot}_{R}}}{s-M_{R}^2+iM_{R}\Gamma^{\rm tot}_{R}}\sqrt{\frac{\Phi(\sqrt{s})}{\Phi(M_{R})}},
\end{equation}
where $M_{R}$, $\Gamma_{R}^{ee}$ and $\Gamma^{\rm tot}_{R}$ are the mass,
partial width to $e^+e^-$ and total width of the assumed resonance $R$.
$\mathcal{B}_{R}$ is the branching fraction for
$R \to\pi^{+}\pi^{-}\eta^\prime$, and $\Phi(\sqrt{s})$
is the two-body PHSP factor of $R \to\rho\eta^\prime$~\cite{PDG}.

In the fit, the correlated and uncorrelated uncertainties
are incorporated and the $\chi^{2}$ is constructed as
\begin{equation} \label{eq:minichifit}
\chi^{2} = \Delta X^{T}\mathcal{M}^{-1}\Delta X,
\end{equation}
where $\Delta X$ is the difference between the measured and predicted
cross sections; the uncertainty of the measured value includes
the uncorrelated statistical and systematic components.
$\mathcal{M}$ is the covariance matrix; its diagonal elements represent the total
uncertainty and off-diagonal elements are correlated systematic
uncertainties.

The systematic uncertainties marked with $^*$ in Table~\ref{tab:sysuncer}
are treated as fully correlated uncertainties, while the other systematic uncertainties
are considered independent for the various c.m.~energies. $\mathcal{M}$ is defined as
\begin{equation} \label{eq:minichifit2}
\mathcal{M}_{i,j} = \sum_{k}x_{i} \; \epsilon_{i,j,k} \; x_{j} \; \epsilon_{j,i,k},
\end{equation}
where $x_{i}$ is the measured value at c.m.~energy $i$, $\epsilon_{i,j,k}=\epsilon_{j,i,k}$
is the common relative systematic uncertainty of $x_{i}$ and $x_{j}$
from correlated source $k$.

Figure~\ref{fig:lineshapefit} shows the result of the fit to the Born cross sections.
There are two solutions with equal fit quality
and very similar mass and width for the resonance,
while the product $\mathcal{B}_{R}\Gamma^{ee}_{R}$ and phases are different in the two solutions.
The goodness of fit is $\chi^{2}$/n.d.f. = 9.4/13 = 0.72, where n.d.f.~is the number of degrees of freedom.
The fit parameters are summarized in Table~\ref{tab:fittingpara}.
Since the mass, width, $n(n')$ and $C_0(C_0')$ values of the two solutions are
consistent within $0.2$ standard deviation, we present the average values of the
two solutions in Table~\ref{tab:fittingpara}.
The statistical significance of this resonance is estimated to be
$6.3\sigma$, by comparing the change of $\chi^{2}$ ($\Delta \chi^{2}=50.5$),
with and without the
$R$ amplitude in the fit and taking the change of degrees of freedom ($\Delta$n.d.f.=4) into account.

\begin{table}[h!]
\begin{center}
\caption{Results of the fit to $e^+e^- \to \eta^\prime\pi^+\pi^-$ Born cross section.
The first uncertainties are statistical and the second systematic. $n'$ and $C_0'$
are parameters of the alternative parameterization of the continuum contribution.}
\label{tab:fittingpara}
\begin{tabular}{lcc} \hline
Parameter                                 & Solution 1       &Solution 2     \\  \hline
$M_{R}$ (MeV/$c^2$)                       & \multicolumn{2}{c}{$2111\pm43\pm25$}   \\
$\Gamma_{R}^{\rm tot}$ (MeV)              & \multicolumn{2}{c}{$135\pm34\pm30$}    \\
$\mathcal{B}_{R}\Gamma^{ee}_{R}$ (eV)     & $0.64\pm0.49\pm0.42$   & $23.3\pm5.3\pm3.3$   \\
$\phi$~(rad)                              & $2.24\pm0.73\pm0.48 $  & $4.46\pm0.06\pm0.10$   \\
$n(n')$                                   & \multicolumn{2}{c}{$4.42\pm0.22\pm0.20~(1.66\pm0.12\pm0.07)$}     \\
$C_0(C_0')$                               & \multicolumn{2}{c}{$921\pm240\pm114~(53.0\pm13.2\pm0.1)$}   \\
\hline
\end{tabular}
\end{center}
\end{table}

The systematic uncertainties of the resonance parameters come from the c.m.
energy calibration, the resonance model, the parameterization of the continuum
and the type of PHSP factor.
The uncertainties of the measured Born cross sections
have been included  in the fit.

The systematic uncertainty of the c.m.~energy is found to be negligible.

To estimate the uncertainty related to the fit model, a modified
Breit-Wigner function, in which the width is energy dependent,
is employed in the fit.
The width of the modified Breit-Wigner function is written as:
\begin{equation} \label{eq:minichifit2}
\Gamma(\sqrt{s})=\Gamma_{R}\left|\frac{p}{p_{R}}\right|^{2 L+1} \frac{M_{R}}{\sqrt{s}} \frac{B(p)}{B\left(p_{R}\right)},
\end{equation}
where $\Gamma_{R}$ is the nominal width. $p$ and $p_{R}$ are the daughter momenta in
the rest frame of $P$, when $P$ taken as $\sqrt{s}$ or $M_{R}$, respectively.
$L$ is the angular momentum of the decay specified in its subscript.
$B(p)$ is the Blatt-Weisskopf form factor~\cite{Blatte}.
The shifts of the mass and width, which are $14$~MeV$/c^2$ and $17$~MeV, respectively,
are taken as the systematic uncertainties.

The uncertainty of the parameterization of the continuum contribution is estimated
by replacing $C_0/s^n$ with an exponential function of the form
$C_0' \cdot e^{-n'(\sqrt{s}-M_{th})}$, where $M_{th}=m_{\rho}+m_{\eta'}$.
The differences of the obtained mass and width,
which are $21$~MeV$/c^2$ and $24$~MeV, respectively, are assigned as the
corresponding systematic uncertainties.

To assess the uncertainty regarding the PHSP factor,
we replace the two-body PHSP factor with an
alternative PHSP factor consist of 90\% two-body and 10\% three-body PHSP factor.
The resulting changes in the fit of $0.4$~MeV$/c^2$ for the mass and $4.1$~MeV for the width
are taken as the systematic uncertainties.

A quadrature sum of all contributions yields total systematic uncertainties
for the mass and width of 25~MeV$/c^2$ and 30~MeV, respectively.

\section{Summary and Discussion}

We present measurements of the Born cross sections for $e^+e^- \to \eta'\pi^+\pi^-$
using the data samples collected by the BESIII detector at c.m.~energies between 2.00 and 3.08~GeV.
The measured Born cross sections are consistent with those of BABAR
but have much improved precision.
The Born cross section line shape fit has two solutions with equal
fit quality and identical mass and width of the resonance, while
the product $\mathcal{B}_{R}\Gamma^{ee}_{R}$ and phase are different in the two solutions.
The statistical significance of the observed resonant structure is $6.3\sigma$,
and its mass, width and $\mathcal{B}_{R}\Gamma^{ee}_{R}$ are determined to be
$M=2111\pm43\pm25$~MeV/$c^2$,$\Gamma=135\pm34\pm30$~MeV and $\mathcal{B}_{R}\Gamma^{ee}_{R}=(0.64\pm0.49\pm0.42)$~eV or $(23.3\pm5.3\pm3.3)$~eV, respectively, where the first
uncertainties are statistical and the second systematic.
The mass and width measured in this work agree with those of
the $Y(2040)$ resonance found in $e^+e^-\to\omega\pi^0$
by BESIII ($M=2034\pm13\pm9$~MeV/$c^2$, $\Gamma=234\pm30\pm25$~MeV)~\cite{Dong2020}
and with those of the $\rho(2150)$ resonance found in $e^+e^- \to \eta'\pi^+\pi^-$
by BABAR ($M=1990\pm80$~MeV/$c^2$, $\Gamma=310\pm140$~MeV)~\cite{BarBarCS} within two standard deviation.

The $e^+ e^- \to \pi^+\pi^-\eta^{\prime}(\eta)$ processes are also
studied in the Resonance Chiral Theory framework and the
extended Nambu-Jona-Lasinio model \cite{Gomez2012,Dai2013,Volkov2014,GammapipiDai,arXiv:2011.09618}. However,
most of the comparisons of experimental data with those theory predictions are performed in
the energy region below 2.0~GeV. With more resonances being included and precise experimental measurements available, these theory models could be tested above 2~GeV in the future.

\section{acknowledgments}

The BESIII collaboration thanks the staff of BEPCII and the IHEP computing center for their strong support. This work is supported in part by National Key Basic Research Program of China under Contract No. 2015CB856700; National key R \& D program of China under Contracts No. 2020YFA0406400 and No. 2020YFA0406300; National Natural Science Foundation of China (NSFC) under Contracts No. 11975118, No. 11625523, No. 11635010, No. 11735014, No. 11822506, No. 11835012, No. 11935015, No. 11935016, No. 11935018 and No. 11961141012, No. 11605196, No. 11605198, No. 11950410506, No. 12061131003, No. 11705192, and No. 12035013;
Natural Science Foundation of Hunan Province under Contract No. 2019JJ30019; The Chinese Academy of Sciences (CAS) Large-Scale Scientific Facility Program; Joint Large-Scale Scientific Facility Funds of the NSFC and CAS under Contracts No. U1732263, No. U1832103, No. U1832207, No. U2032111, No. U2032105; CAS Key Research Program of Frontier Sciences under Contracts Nos. QYZDJ-SSW-SLH003, QYZDJ-SSW-SLH040; 100 Talents Program of CAS; INPAC and Shanghai Key Laboratory for Particle Physics and Cosmology; ERC under Contract No. 758462; German Research Foundation DFG under Contracts Nos. 443159800, Collaborative Research Center CRC 1044, FOR 2359, FOR 2359, GRK 214; Istituto Nazionale di Fisica Nucleare, Italy; Ministry of Development of Turkey under Contract No. DPT2006K-120470; National Science and Technology fund; Olle Engkvist Foundation under Contract No. 200-0605; STFC (United Kingdom); The Knut and Alice Wallenberg Foundation (Sweden) under Contract No. 2016.0157; The Royal Society, UK under Contracts Nos. DH140054, DH160214; The Swedish Research Council; U. S. Department of Energy under Contracts Nos. DE-FG02-05ER41374, DE-SC-0012069


\end{document}